## Study of the Cataclysmic Variable 1RXS J174320.1-042953

D. V. Denisenko[1], F. Martinelli[2]

[1] Sternberg Astronomical Institute at Lomonosov Moscow State University, Russia; e-mail: d.v.denisenko@gmail.com
[2] Montecatini Astronomical Centre, Italy; e-mail: info@astronomicalcentre.org

We report on the photometric analysis of the poorly studied cataclysmic variable in Ophiuchus 1RXS J174320.1-042953 (DDE 11). Results of the monitoring with the Bradford Robotic Telescope (BRT) are presented, as well as the time-resolved photometry obtained at Montecatini Astronomical Centre. The long-term behavior of J1743-0429 shows it is a magnetic cataclysmic variable (polar) with the large difference between the low and high states. The orbital period obtained from our observations is 0.0866d, or 2.08hr, close to the lower boundary of the period gap of cataclysmic variables.

The cataclysmic variable DDE 11 = 1RXS J174320.1-042953 was identified by Denisenko and Sokolovsky (2011) from the magnitudes in USNO-B1.0 catalogue ($B1$=16.12, $R1$=15.15, $B2$=18.95, $R2$=18.13). They suggested the star to be a dwarf nova which was in outburst on July 01, 1954 when the 1$^{st}$ epoch Palomar Observatory Sky Survey blue and red plates were obtained.

After the discovery of variability on the digitized photographic plates and in USNO-B1.0, one of us (D.D.) has started monitoring the variable with the Bradford Robotic Telescope located on Tenerife (Canary Islands, Spain). The images from the 0.35-m f/5.3 telescope with FLI Microline CCD (1k × 1k pixels) were requested via the website http://www.telescope.org. Due to the large number of observers using this publicly available facility the requests are fulfilled at irregular intervals of time. Typically one has to wait for the image from 2 weeks to 2 months after submitting the request. However, even this cadence of observations allows one to obtain the valuable data on the long term behavior of cataclysmic variables.

In the case of 1RXS J174320.1-042953 (hereafter referred to as J1743-0429 for brevity) the observations with BRT have shown that the initial hypothesis of this star being a dwarf nova was incorrect. Instead of spending most time at quiescence with occasional outbursts, J1743-0429 was found to be always near the high state, yet slowly growing with time. The results of BRT observations are given in Table 1.

Table 1: BRT observations of J1743-0429.

| Date | JD | Mag |
|---|---|---|
| 2011 May 22 | 2455703.627 | 16.14 |
| 2011 Aug. 04 | 2455777.509 | 15.88 |
| 2012 July 16 | 2456124.560 | 15.58 |

Observations were performed in the white light with 60-sec exposures. Magnitudes were measured using USNO-A2.0 0825-11051991 (R.A.=17 43 22.71, Dec.=-04 30 29.6, $R$=15.7) as a reference star and USNO-A2.0 0825-11051825 (R.A.=17 43 22.49, Dec.=-04 30 59.8, $R$=16.3) as a check star. The variable, reference and check stars are shown in Fig. 1 which is a sum of two best BRT exposures. It should be noted that the image scale of Galaxy camera is 1.44″/pixel, and there is another star only 6″ to the South of the variable. This is why the sizes of aperture, gap and annulus rings were all set equal to 3 pixels.

Following the request by D.D., time-resolved photometry of J1743-0429 was performed by F.M. at Montecatini Astronomical Centre (http://www.astronomicalcentre.org) on two nights using the 0.35-m telescope with ST8-XME CCD. 70-sec unfiltered exposures were taken on both nights. 4.6-hr long time series was obtained on 2012 July 16/17 with yet 3.5 hours on July 18/19. The images were calibrated using flat fields and dark frames obtained on the same night. The photometry was made using the same reference and check stars as indicated above.

The light curves for both nights are shown in Fig. 2. The combined data set was analyzed using the WinEffect software by Dr. V. P. Goranskij. Period search was performed by Deeming and Lafler-Kinman methods. Both algorithms have provided the best period value equal to $0^d.0866(3)$, or 11.55 cycles per day. The periodogram in the range 8-16 cycles/day obtained by Lafler-Kinman method is shown in Fig. 3, and the resulting phased light curve - in Fig. 4.

The main features of the light curve are its quite large amplitude ($0.8^m$) and the dip by about $0.4^m$ shortly before the maximum. Together with the orbital period close to the bottom edge of CV period gap these features are indicating to the polar nature of this cataclysmic variable.

One can also suggest a hint of the shorter period variations of light superimposed on the main orbital curve. It is possible that this CV is an intermediate polar with a quickly rotating white dwarf. Photometric time series with a larger aperture telescope are required to study this variable in more detail. We also encourage the spectroscopic study to measure the relative intensities of He II 4686 Å and $H_\beta$ lines.

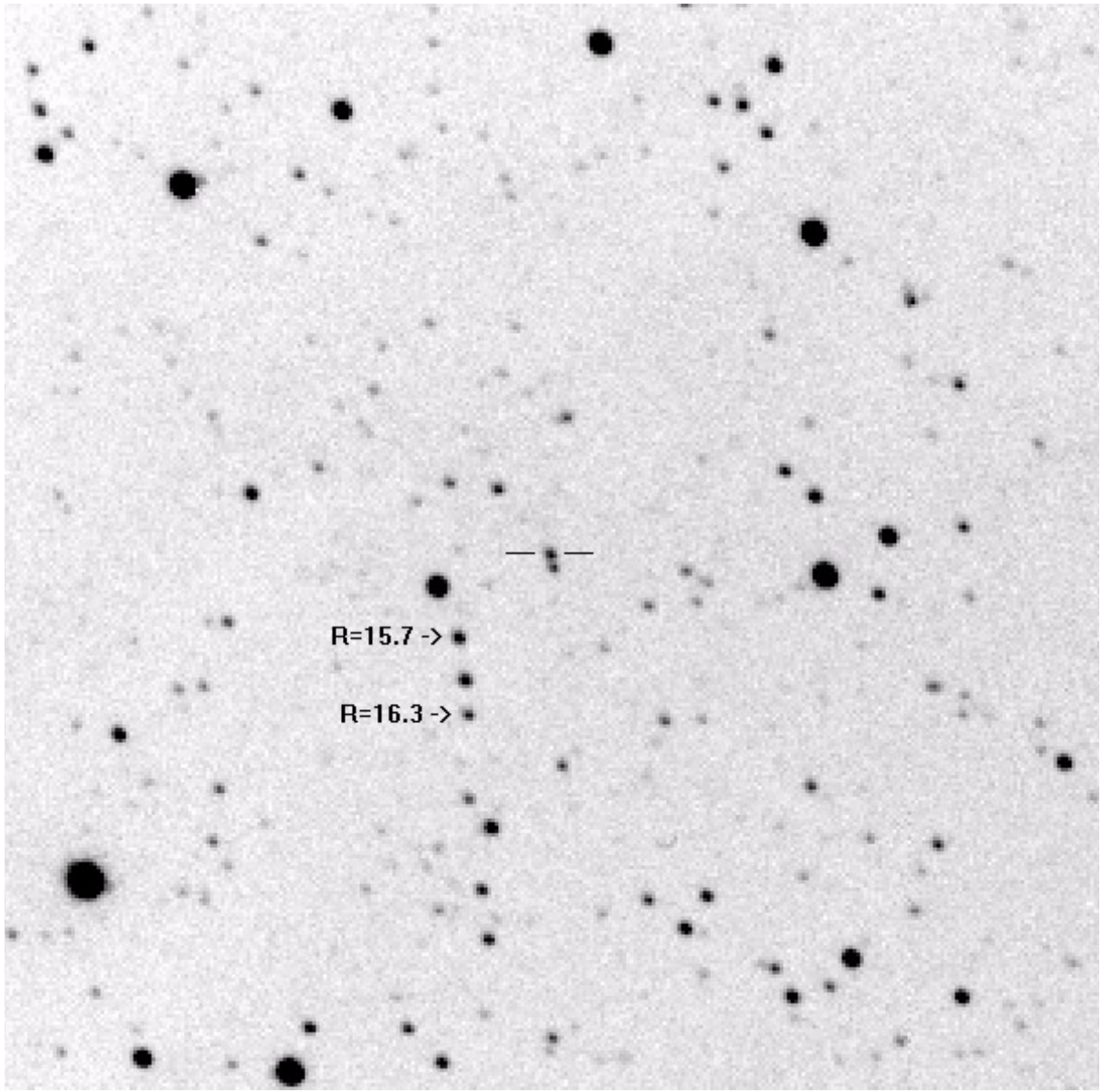

**Figure 1.** Finder chart of J1743-0429 from the Bradford Robotic Telescope image. North is up, East is to the left. Field of view is 14′×14′. The variable is marked with dashes, the reference and check stars are shown by arrows together with their USNO-A2.0 *R* magnitudes.

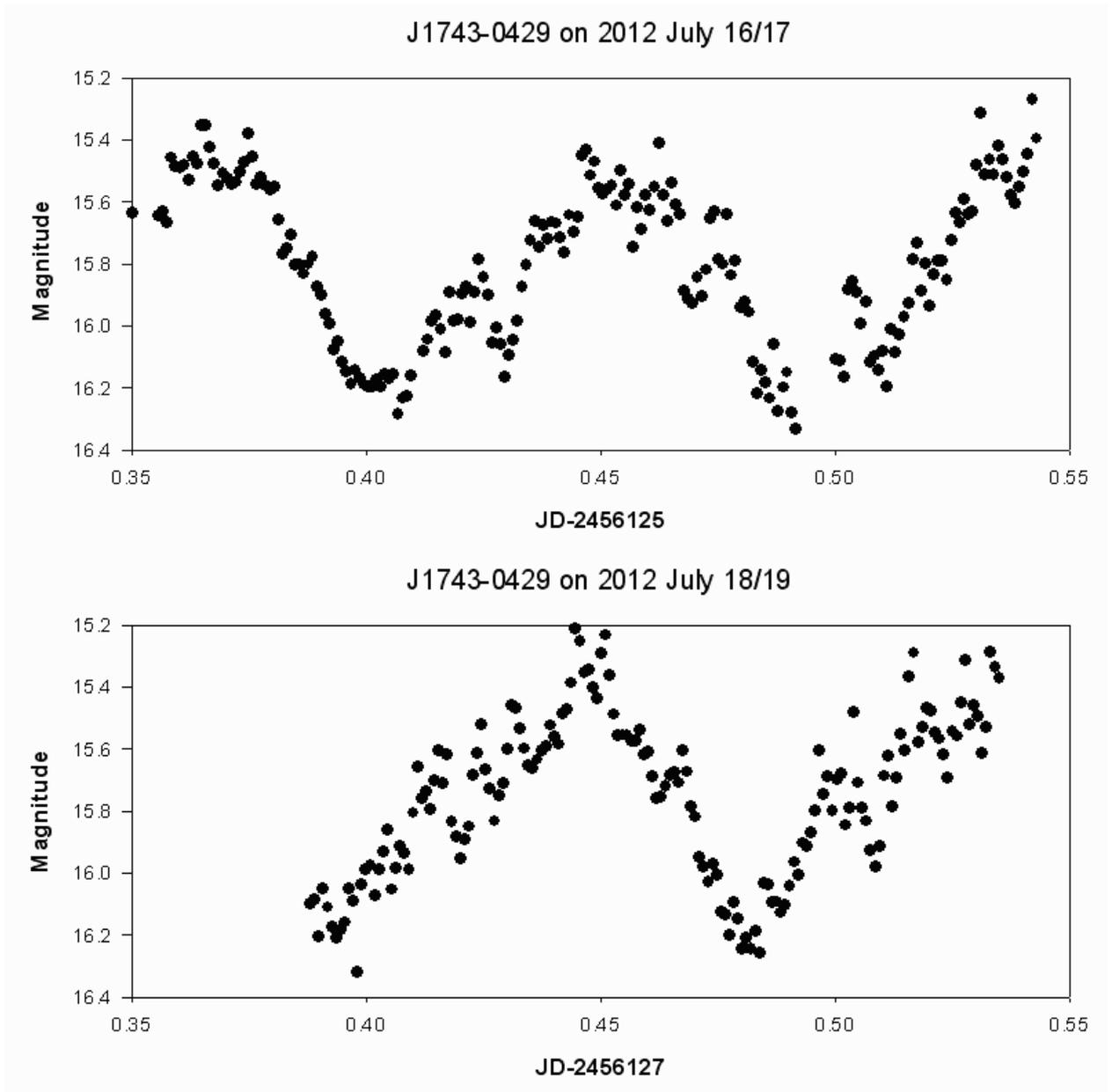

**Figure 2.** Light curves of J1743-0429 obtained at Montecatini Astronomical Centre on 2012 July 16/17 (top) and 2012 July 18/19 (bottom).

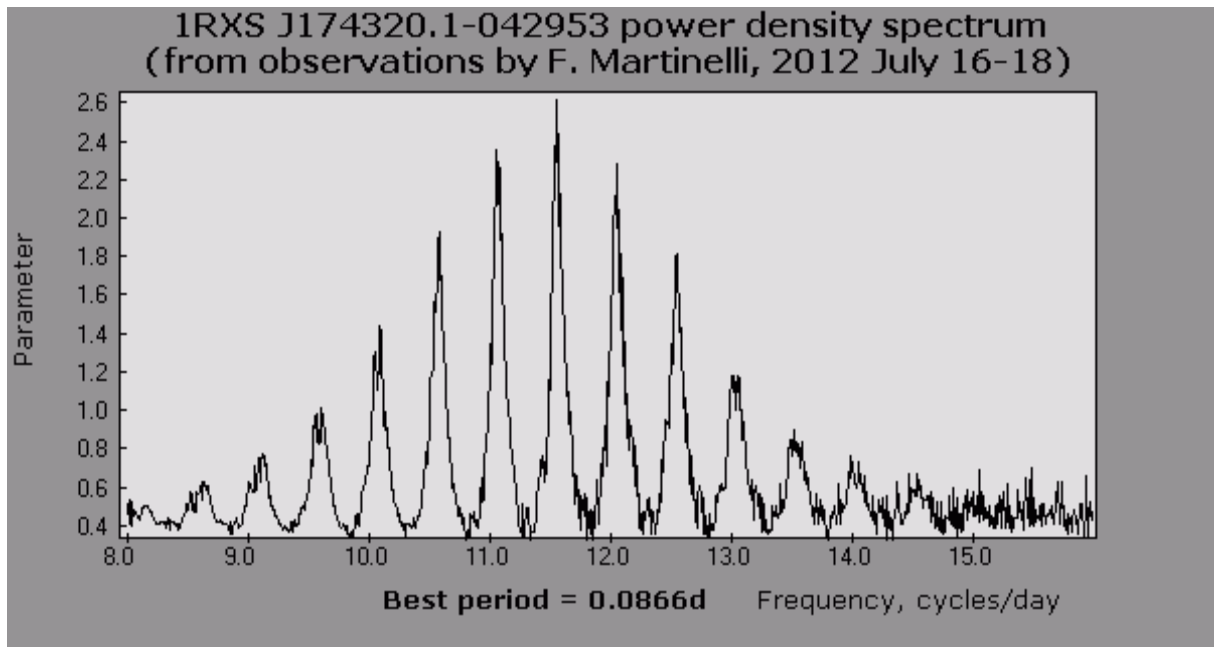

**Figure 3.** Periodogram of J1743-0429 observations obtained with the Lafler-Kinman method.

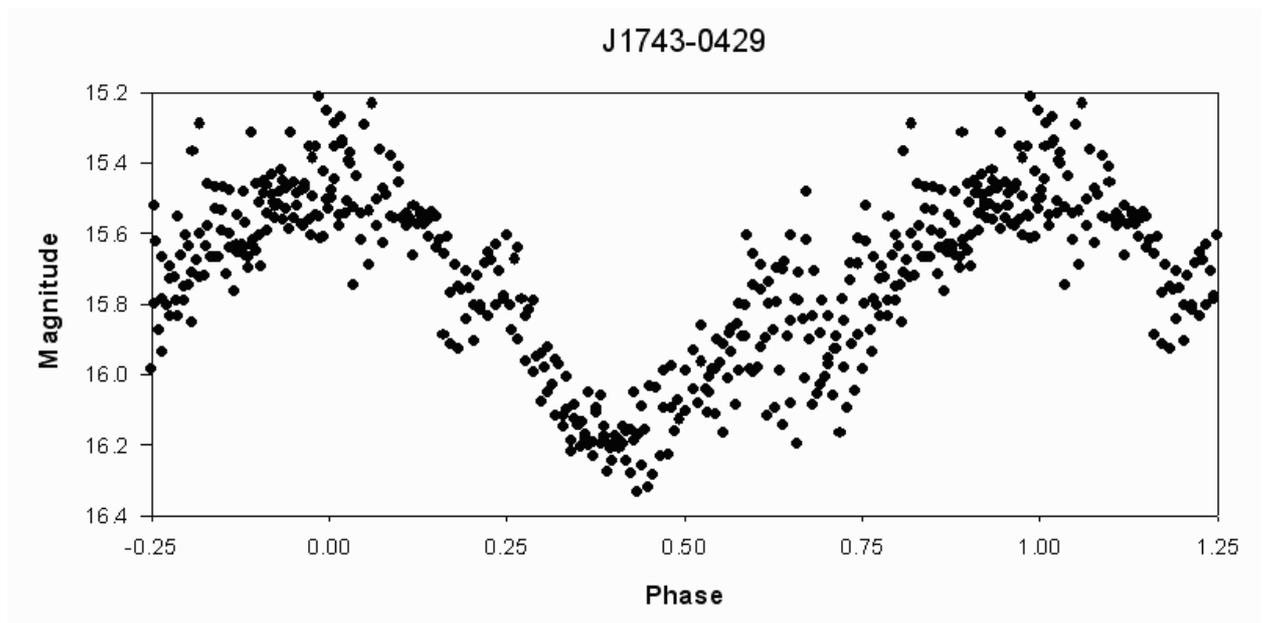

**Figure 4.** Light curve of J1743-0429 folded at the best period 0.0866d.